\documentclass[a4paper]{aa} 
\usepackage{graphicx}
\usepackage{natbib}
\bibpunct{(}{)}{;}{a}{}{,} 
\usepackage{rotate}
\usepackage{txfonts}

\begin{document}

\title{MCAO near--IR photometry of the Globular Cluster NGC~6388:\\
MAD observations in crowded fields 
\thanks{Based on observations collected at the European Southern
  Observatory, Chile, as part of MAD Guaranteed Time Observations, and
on  NASA/ESA Hubble Space Telescope observations (GO-10775).}}
   \author{A. Moretti
   		\inst{1}
             \and
	     G. Piotto
              \inst{2}
            \and
	    C. Arcidiacono
           \inst{1}
             \and
	     A.~P. Milone
              \inst{2}
	   \and
	   R. Ragazzoni
           \inst{1}
	   \and
	   R. Falomo
           \inst{1}
	   \and
	   J. Farinato
           \inst{1}   
	   \and
	   L.~R. Bedin
	   \inst{3}
	   \and
	   J. Anderson
	   \inst{3}
	   \and
	   A. Sarajedini
	   \inst{4}
	   \and
	   A. Baruffolo
	   \inst{1}
	   \and
	   E. Diolaiti
	   \inst{5}
	   \and
	   M. Lombini
	   \inst{5}
	   \and
	   R. Brast
	   \inst{6}
	   \and
	   R. Donaldson
	   \inst{6}
	   \and
	   J. Kolb
	   \inst{6}
	   \and
	   E. Marchetti
	   \inst{6}
	   \and
	   S. Tordo
	   \inst{6}
          }


   \institute{INAF - Osservatorio Astronomico di Padova,
     Vicolo dell'Osservatorio 5, I-35122 Padova                    
     \and
     Dipartimento di Astronomia - Universit\`a di Padova,
     Vicolo dell'Osservatorio 3, I-35122 Padova
     \and
     Space Telescope Science Institute, 
     3800 San Martin Drive, Baltimore, MD 21218
     \and
     Department of Astronomy, University of Florida, 
     Gainesville, FL 32611, USA
     \and
     INAF - Osservatorio Astronomico di Bologna, 
     via Ranzani 1, I-40127 Bologna
     \and
     European Southern Observatory, 
     Karl-Schwarzschild-Strasse 2, G-85748 Garching
   }

   \date{6 Oct 2008}

\abstract{Deep photometry of crowded fields, such as Galactic Globular
  Clusters, is severely limited by the actual resolution of
  ground--based telescopes. On the other hand, the Hubble Space
  Telescope does not provide the near--infrared (NIR) filters needed
  to allow large color baselines.}{In this work we aim at
  demonstrating how ground based observations can reach the required
  resolution when using Multi-Conjugated Adaptive Optic (MCAO) devices
  in the NIR, such as the experimental infrared camera (MAD) available
  on the VLT. This is particularly important since these corrections
  are planned to be available on all ground--based telescopes in the
  near future.}{ We do this by combining the infrared photometry
  obtained by MAD/VLT with ACS/HST optical photometry of our
  scientific target, the bulge globular cluster NGC~6388, in which we
  imaged two fields. In particular, we constructed color-magnitude
  diagrams with an extremely wide color baseline in order to
  investigate the presence of multiple stellar populations in this
  cluster.} { From the analysis of the external field, observed with
  better seeing conditions, we derived the deepest optical-NIR CMD of
  NGC~6388 to date.  The high-precision photometry reveals that two
  distinct sub-giant branches are clearly present in this cluster.  We
  also use the CMD from the central region to estimate the distance
  [$(m-M)_\circ=15.33$] and the reddening ($E(B-V)=0.38$) for this
  cluster.  We estimate the age to be ($\sim11.5\pm 1.5$Gyr).  The
  large relative-age error reflects the bimodal distribution of the
  SGB stars.}{ This study clearly demonstrates how MCAO correction in
  the NIR bands implemented on ground based telescopes can complement
  the high-resolution optical data from HST.}

\authorrunning{} 
\titlerunning{MAD performances in NGC~6388} 
\maketitle

\section{Introduction}\label{sec:intro}
NGC~6388 and NGC~6441 are among the most intriguing Galactic Globular 
Clusters (GCs).  They are both located in the Galactic bulge.  Early 
studies of their integrated light by \citet{rich+1993} revealed an 
anomalous UV-excess.
Both clusters are metal--rich ([Fe/H]$\sim -0.6$ and $\sim -0.5$, 
respectively) and should therefore possess a red horizontal branch (HB), 
with no hot blue HB stars.  However, deep high-resolution images obtained 
by the Hubble Space Telescope (HST) revealed that both clusters exhibit 
a very long tail of blue HB stars that explains the UV excess 
\citep{rich+1997}.  In addition, the HBs of NGC~6388
and NGC~6441 show a sloped appearance \citep{raim+2002,buss+2007}, which can
not be reproduced by stellar evolution models with the 
clusters' known metallicities and typical GC abundance ratios
\citep[see also][for a detailed discussion]{cate+2006}.  
When compared with 47 Tuc, NGC~6388 shows a broader color--magnitude
diagram (CMD), both in the turnoff region and along the Red Giant Branch
(RGB), which is at least in part attributed to the presence of
differential reddening \citep{buss+2007}.

What is then responsible for the long blue tail of hot stars in these
clusters?  The solution to this problem, which might be the most
extreme manifestation of the long-standing second-parameter problem,
has yet to come.  One resolution might be related to the observation that
the most massive GCs of our Galaxy ($\omega$Cen, NGC~2808, NGC~1851)
possess CMDs which split into multiple evolutionary sequences, either in
the MS region \citep[NGC~2808,][]{piot+2007}, or in the Subgiant
Branch region \citep[NGC~1851,][]{milo+2008}, or in the MS, SGB, and RGB
regions \citep[$\omega$Cen,][]{lee+1999,panc+2000,bedi+2004,vill+2007}.

The recent evidence \citep{piot2008} that the SGB of NGC~6388 is
split into two distinct branches gives further support to the
possibility that this cluster may also host more than one stellar
population.

\citet{ree+2002} demonstrated that the peculiarities of NGC~6388's 
CMD (i.e. the broad RGB and the bimodal HB) can be explained
by allowing a double generation of stars with a mild age-metallicity
relation.  Alternatively, the second generation of stars could be formed
from material already enriched in CNO--cycle products by AGB stars, and
could therefore have enhanced helium abundance
\citep{danto+2002,danto+calo2004}.
We note that for helium to explain the blue tail, it cannot involve simple 
deep mixing, but must involve an actual abundance variation.  Such a 
helium-enhanced population could have an age difference from the first
generation of only a few hundred million years, not measurable on the CMD.
\citet{yoon+2008} simulations of the observed CMDs of NGC~6388 and NGC~6441
confirm that the only mechanism able to produce a tilted HB and the
observed RR Lyr properties is a self-enrichment scenario, 
however, they find a lower helium content for the enhanced population
than do \citet{calo+danto2007}.
Spectroscopic analysis by \citet{carr+2007} confirms that the BHB can be
caused by this self-enriched helium-enhancement; they also rule out large
age differences between the generations.

A strong indicator of the presence of multiple generation of stars
with different helium content in NGC~6388 would be the discovery of
the presence of multiple main-sequence populations \citep[see][]{piot+2007}. 
However, since the cluster is located in the Galactic bulge, the strong
contamination by field stars and the differential reddening make this
investigation rather complicated.

In principle, near--infrared (NIR) observations should improve our
ability to distinguish different stellar generations in NGC~6388,
because of the reduced effect of differential reddening and because of
the larger color baseline obtained when coupling the NIR observations
with optical ones. Theoretical models predict, for example, that
sequences characterized by a different helium content, would appear
well separated, and this is easier to distinguish with a larger color
baseline. In the case of NGC~6388 a difference of 0.15 in helium
content \citep[as suggested by][]{danto+calo2008} would cause a color
split of 0.13 mag in $m_{F606W}-K_s$ which corresponds to 0.04 in
$(B-V)$.  On the other hand the broadening effect due to the effects
of differential reddening can be minimized as in
\citet{sara+2007} and \citet{milo+2008}, thus making very appealing this
combination of passbands.  In addition to exploring the main
sequence, the NIR observations would also allow us to verify that the
presence of the double SGB identified by \citet{piot2008} is not an
artifact of differential reddening.  Moreover, theoretical
isochrones show that the chosen optical--NIR color is much more
sensitive to age differences (1 Gyr corresponding to $m_{\rm
F606W}-K_s \sim 0.03$ and $B-V \sim 0.01$) with respect to pure
optical colors.

For all the above reasons, we decided to observe this massive bulge GC 
using the Multi--conjugated Adaptive optics Demonstrator (MAD) on the
VLT in hopes of obtaining accurate photometry in order to explore the 
evidence for multiple populations in this cluster.  With this goal in 
mind, we also tried to test MAD capabilities under two different conditions.
We observed two fields in NGC~6388:  one located at the center of the 
cluster, where we can sample a large number of stars, though crowding 
severely affects the images (NGC~6388-a), and a second less dense field 
(NGC~6388-b), located 110 arcseconds away from the center.  The external 
field has been used as a test of MAD performance in terms of high-precision
photometry.

\section{Observations and analysis}\label{sec:obs}
\begin{figure*}[htbp]
\begin{center}
\includegraphics[width=0.44\textwidth]{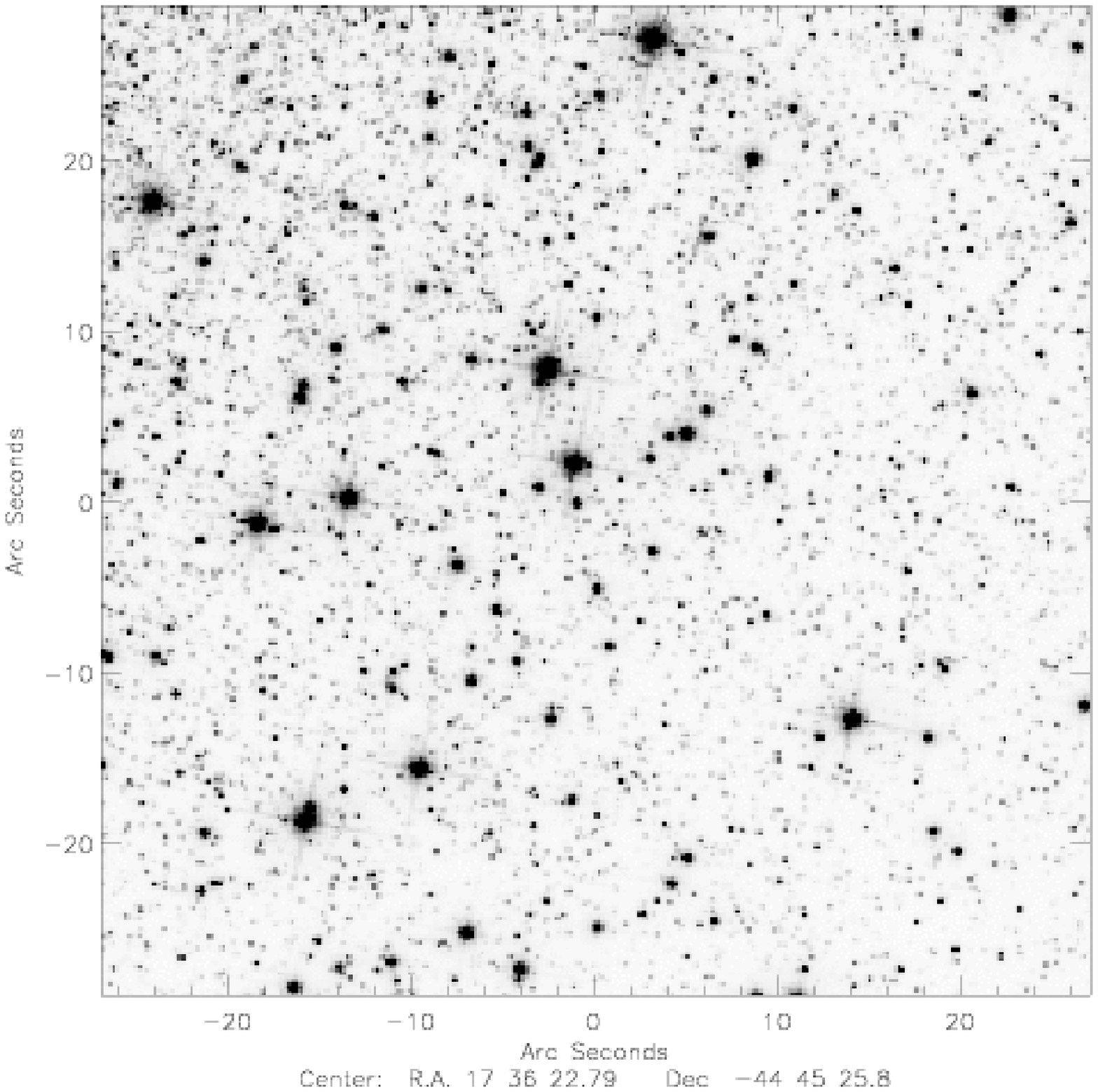}
\includegraphics[width=0.44\textwidth]{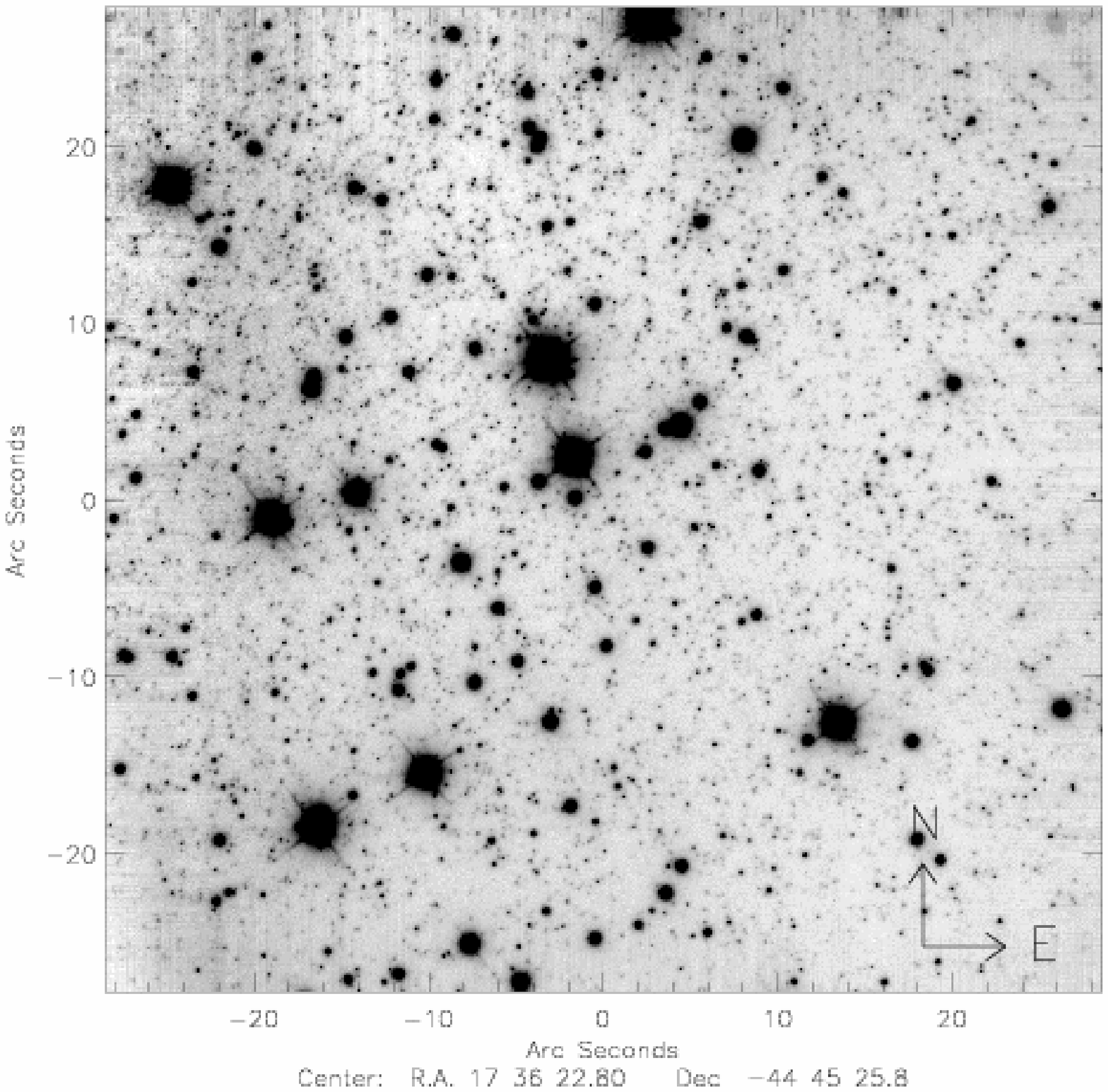}
\caption{The external field, NGC~6388~b, imaged by ACS/HST in the
  F814W filter (left) and by MAD/VLT (right). Exposure times are 1790s
  and 240s, respectively.}
\label{fig:mad_fov}
\end{center}
\end{figure*}

Near-IR observations were carried out with the Multi-Conjugated
Adaptive Optics Demonstrator (MAD) mounted at the VLT \citep{marc+2007}.
MAD is a prototype Multi Conjugated Adaptive Optics (MCAO) system which 
has been designed to test in the laboratory and on sky the feasibility of
different MCAO image reconstruction and correction techniques, in view
of future applications with the 2nd Generation VLT and
E-ELT instrumentation.

The MAD imager is a 1 arcmin field-of-view (FoV) infrared camera 
scanning 2 arcmin FoV (round). The CCD has 2048 $\times$ 2048 pixels 
with a scale of 0.028 arcsec/pixel \citep{marc+2006}.

The wavefront sensor that we used is a Layer Oriented (LO) multi
pyramid sensor
\citep{raga1996,raga+2000,vern+2005,arci+2006,arci+2007} that can use
from 3 to 8 ``faint'' ($V<18$) stars, for which the integrated light
reaches $V=13$.  The stars can be placed anywhere in the telescope
FoV. The LO technique has a great advantage, with respect to the Star
Oriented (SO) one, because it offers the possibility of using fainter
stars \citep[see][for a detailed discussion]{raga+fari1999,ghed+2003}.

Observations of the two fields in NGC~6388 were made in September 2007
using the full MCAO capability. 
These were part of a set of science observations aimed at verifying 
the capability of MAD at VLT \citep{gull+2008,falo+2008}.
In Table \ref{tab:log} we show the
observation log: the first field (labeled NGC~6388-a) is located at
the center of the cluster (17:36:17.86, -44:44:05.60), and the second
one (NGC~6388-b) at RA(J2000.0)=17:36:22.86, DEC(J2000.0)=-44:45:35.53 (see
Fig.\ref{fig:mad_fov}).  The last column indicates the $V$ magnitude of
the stars used to perform the AO correction.

It is remarkable that the reference stars used for the MCAO correction
in the case of NGC~6388-b are significantly fainter than the ones for
NGC~6388-a and, in general, much fainter than the ones usually adopted
for the Star Oriented Wavefront Sensor. In spite of this, the outcome
is significantly better, as the initial seeing is significantly
better, although well within the median seeing of the VLT site.

In both cases, dedicated sky exposures have been taken at an
appropriate distance from the scientific frame. For these images the
background subtraction has been performed taking into account the sky
level and shape determined from the offset frames.

\begin{table*}[htbp]
\caption{Observations log}
\label{tab:log}
\centering
\begin{tabular}{|l|c|c|c|c|c|c|c|c|}
\hline\hline 
Name & Night & Filter & DIT & NDIT & T$_{exp}$ &$<FWHM>$& Seeing &
$m_{\rm F606W}$ (AO stars) \\
\hline						  	
NGC~6388-a & 26/9/2007  &  $K_s$  & 10 &  4 & 3000 &  0.27  & 1.76  &
13.6 - 13.9 - 14.0 - 14.8 - 15.1   \\ 
NGC~6388-b & 27/9/2007  &  $K_s$  & 10 & 24 & 5040 &  0.15  & 0.46  &
15.0 (2) - 15.6 - 15.7 - 16.3   \\ 
\hline	
\end{tabular}
\end{table*}

The data-reduction process started with the trimming of the images in
order to exclude from further analysis the overscan region.
Flat--field images have been taken on sky at the beginning of each
night in the $K_s$ filter (centered at 2.15 $\mu$).
After excluding the bad flats we finally combined them into a
median frame that has been used to correct the scientific frames.
Flat--field images have been used to produce the appropriate Bad Pixel
Mask as well.

Each scientific frame was then corrected with the flat--field, and bad
pixels have been removed using the constructed mask.
After having verified that the sky variations during the observations
were negligible, we decided to use one single sky image to correct all
scientific frames.

In the case of crowded fields (such as globular clusters) it is normal
practice to use offset positions to determine the mean level and the
shape of the sky/background correction by making the median of these
sky frames. In particular, when possible we used the ones closer in
time to the scientific images. The final sky/background frame has been
normalized to the median counts of the scientific frame before
being subtracted.

\begin{figure}[htbp]
\begin{center}
\includegraphics[width=0.45\textwidth]{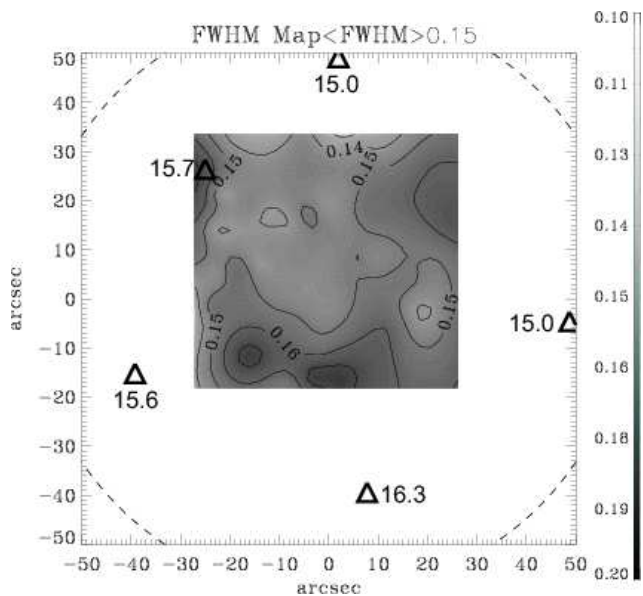}
\caption{FWHM map of the field of NGC~6388-b that was observed with
  MAD. Triangles represent the stars used to perform the AO
  correction. Contours are drawn and labelled according to the FWHM
  measured on observed stars. Magnitudes for AO stars are given in
  $m_{\rm F606W}$.}
\label{fig:fwhm_maps}
\end{center}
\end{figure}

We finally measured the overall stellar FWHM over the FoV, to check
the performance of the AO correction.  Fig.\ref{fig:fwhm_maps} shows,
as an example, the FWHM map obtained for the field NGC~6388-b.  Note
the strong improvement of the seeing due to the AO correction (Column
seven of Table \ref{tab:log}) with respect to the ESO DIMM monitor
seeing (Column eight of Table \ref{tab:log}).

The standard star  FS5, taken from UKIRT Faint Standard catalogue
\citep{ukirt_std}, has been observed, and then used to calibrate
the photometry.  Unfortunately, the observation timing and the need to
manually operate the filters did not allow us to take frames in other
filters (such as $J$ or $H$), so that no color dependence of the 
photometric zeropoint could be determined. 

Our dataset in the central region consists of 75 images with DIT=10 s,
NDIT=4 that are characterized by an average FWHM of 0.27 arcsec.
As for the external field, we have 21 images with DIT=10 s, NDIT=24,
characterized by a much smaller stellar FWHM (on average 0.15 arcsec).

\subsection{Photometry}
We adopted two different approaches for the photometric analysis of
our observations. One 
consisted of making an average image out of the
observed ones (excluding the ones with the worst seeing), and
performing photometry with DAOPHOT/ALLSTAR \citep{daophot} on this image. 
This approach was used for the central field, where we selected
the exposures with a stellar FWHM smaller than 0.5 arcsec (25 out of 75), 
and combined them into a single one with a total exposure time of 1000 s.

The second approach consists of extracting the PSF on each image,
registering it to a common reference frame and then use ALLFRAME to
measure stellar magnitudes on each single frame.
In this case, we used the six best images (those with FWHM smaller than
0.13 arcsec) for a total exposure time of 1440 sec. 
We adopted this approach
for the analysis of the external field (NGC~6388-b).

In both cases we find that the stellar PSF is better modeled with a
quadratic dependence on the scientific frame, even though it is fairly
uniform on the frame (up to a few percent), due to the excellent
Adaptive Optics correction.  We also evaluated all the possible
expressions for the analytic part of the PSF, and selected the
one giving the lowest chi-squared.  The analytic profile which best
fits the stellar PSF turned out to be a Penny function, i.e. a profile
which has both a Gaussian core and power--law wings \citep{penn1986}. 

\section{Results}\label{sec:results}
As we discussed above, MAD observations were available only in the
$K_s$ band. In order to construct a CMD and ensure the largest
possible color baseline, we used the F606W ACS/HST photometry
\citep{ande+2008} from the GO-10775 Treasury project
(P.I. A. Sarajedini). The ACS images cover both MAD fields. The
photometric catalogue has been produced by \citet{ande+2008} using a
dedicated procedure developed over the last few years \citep[see
also][]{ande+king2000,ande+king2006} and placed onto the VEGAMAG system
following \citet{bedi+2005} using the zeropoints of
\citet{siri+2005}. We also corrected the catalogs for the
presence of differential reddening and possible zeropoint differences,
using the empirical corrections described in \citet{sara+2007,milo+2008}.

Below, we will use observations of the external field of NGC~6388 to
investigate the possible presence of multiple main sequences and/or
SGBs. Because of the lower crowding, these images have a better
photometric quality with respect to the central field ones, and a high
level of photometric precision is needed to disentangle multiple
populations in the CMD.  We will use the central, more populous field
to study the RGB and the HB morphology.

\subsection{External field (NGC~6388-b)}\label{sec:ext}
In addition to being less crowded, the images of the external field
also have better seeing.  In particular, 6 out of 21 frames are
characterized by a stellar FWHM of less than 0.13 arcsec.  In order to
maximize the precision of our photometry, we used these 6 best seeing
images, for a total exposure time of 24 minutes.  (We verified this
decision by adding the remaining images, and noticing a degradation in
the photometric quality.)

\begin{figure}[htbp]
\begin{center}
\includegraphics[width=0.45\textwidth]{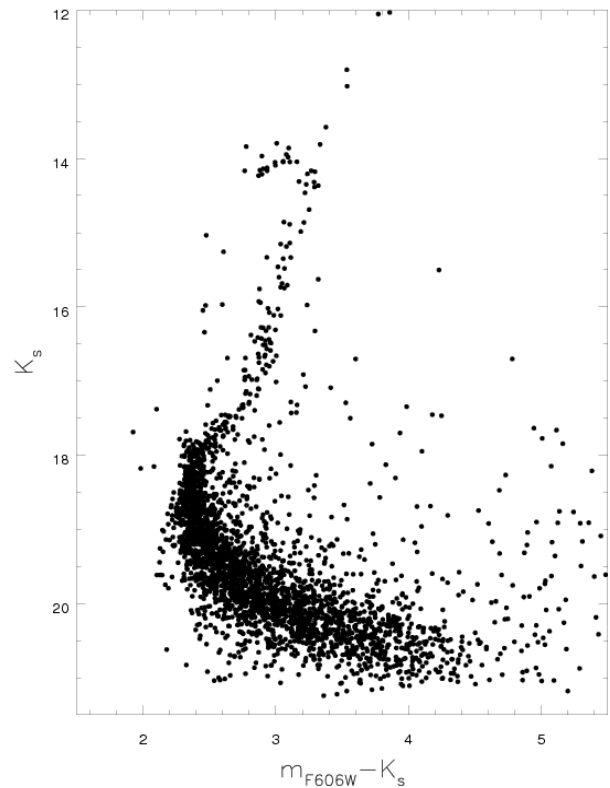}
\caption{Color--magnitude diagram of the external region of
  NGC~6388.}
\label{fig:cmd_6388}
\end{center}
\end{figure}

We constructed the optical--NIR color--magnitude diagram by matching
our stars with the ones measured on the ACS field.  We obtained 7162
matches, and further selected them according to the parameters given
by DAOPHOT, i.e. we excluded stars with a formal photometric error
greater than 0.1 and with a $\chi^2$ greater than 5. We also imposed
the requirement that our matched stars have the same centroid within
$\pm 0.028$ arcsec in both optical and NIR images.  We excluded from
the following analysis stars found outside a circle having a radius of
800 pixels from the center of the image.  This choice is motivated by
the fact that the outermost region of the frame could be deteriorated
because of non-optimal sky subtraction and, moreover, the outer
regions could also be affected by imperfect distortion corrections.
Therefore, the covered field is about $48 \times 48$ arcsec$^2$.  A
total number of 2958 stars passed all the selection criteria.  The
optical--NIR CMD for them is shown in Fig.\ref{fig:cmd_6388}.

This is the deepest CMD available for this cluster in the NIR bands,
and demonstrates that a ground-based 8-m class telescope, equipped
with the appropriate AO correction, can give performance comparable to
the ones reached by HST, thus nicely complementing the information at
optical wavelengths.

\begin{figure*}[htbp]
\begin{center}
\includegraphics[width=0.75\textwidth,angle=270]{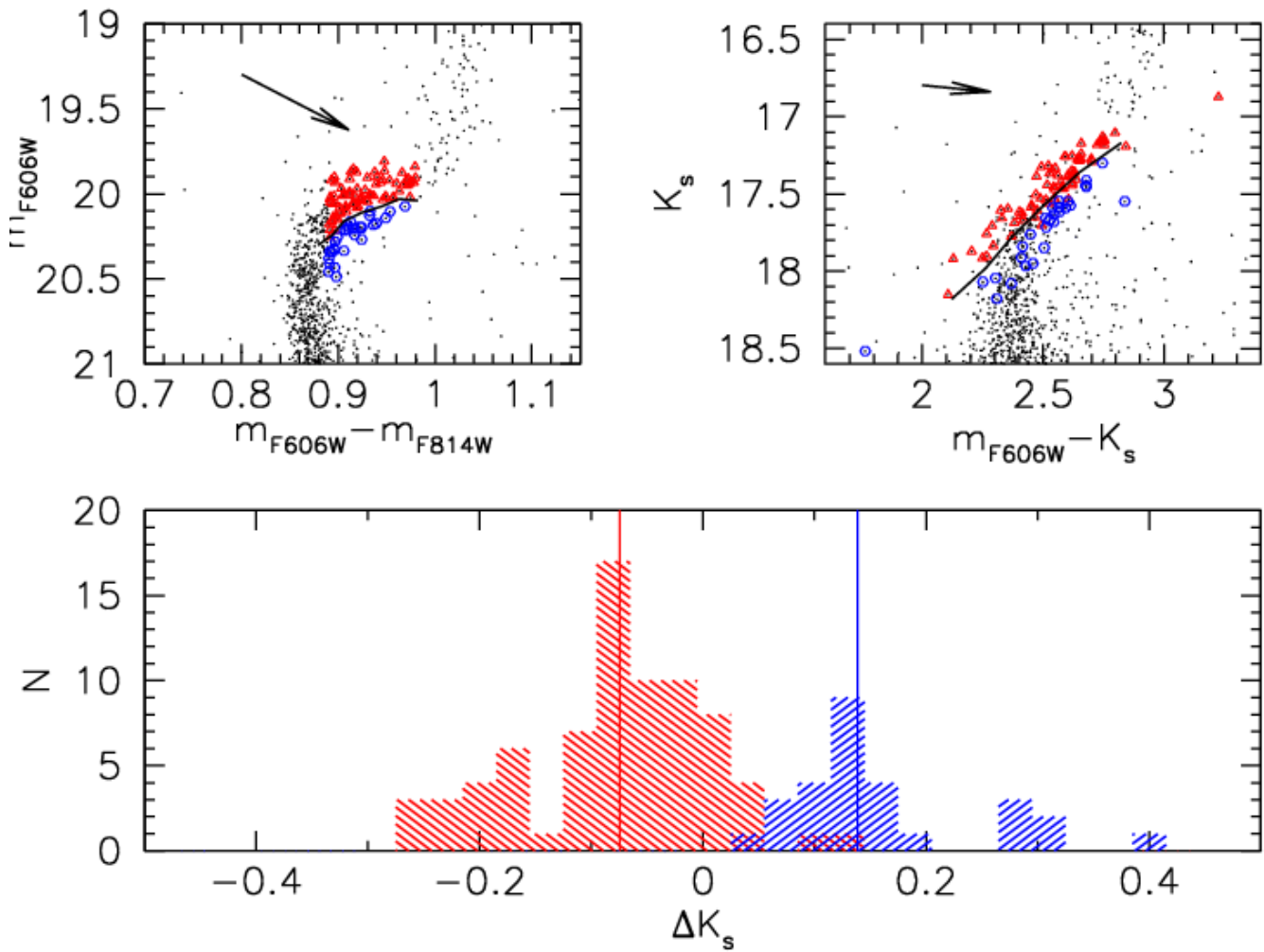}
\caption{Color--magnitude diagrams of the external region of
  NGC~6388. Upper left panel shows the pure optical CMD, with magnitudes
  coming from ACS/HST photometry, for the stars located in the MAD
  field: red triangles highlight stars on the bright SGB, blue open
  circles shows the fainter SGB stars. 
  Upper right panel shows the optical-NIR CMD. Red and blue dots
  represent the same stars identified in the upper left panel CMD. 
  Lower panel shows the magnitude distribution of
  the two subpopulations with respect to a fiducial line drawn in the
  optical--NIR CMD.}
\label{fig:cmd_seq}
\end{center}
\end{figure*}

\citet[][see his Fig. 6]{piot2008} has shown  
that the SGB in the ACS CMD of NGC~6388 is divided into two branches, 
very similar
to what has been identified in the SGB of NGC~1851 \citep{milo+2008}.
In top-left panel of Fig.\ref{fig:cmd_seq} we show ACS/WFC photometry
from \citet{piot2008}, on which we drow by-eye a fiducial line that
clearly separates the two SGB populations. The stars with a color
between ($(m_{\rm F606W}-m_{\rm F814W}$) $=$0.89 and 0.98 , and
brighter than the fiducial-line, are highlighted with red triangle
symbols (hereafter, according to \citet{milo+2008}, notation we will
refer to these stars as bright SGB, bSGB). The stars in the same
color-interval which fell below the same fiducial line are marked with
blue circles (faint SGB, fSGB).

In the top-right panel of Fig.\ref{fig:cmd_seq} we show --for the same
stars-- the color-magnitude obtained combining MAD observations in
$K_s$ with ACS/WFC photometry in filter F606W.

When we plot the stars classified as bSGB and fSGB
with the same symbols in this optical-NIR CMD, we
can clearly see that bSGB are also systematically
brighter than fSGB stars also in filter $K_s$. 

This is of particular relevance, since 
the absorption coefficient in $K_s$ band is 13\% of the one present in
the visible band (here F606W), i.e. $0.37 \times E(B-V)$
\citep{schl+1998} instead of $2.89 \times E(B-V)$
\citep{schl+1998,bedi+2005,siri+2005,gira+2008}.
Furthermore, we recall that we corrected the catalogs for
  differential reddening, as in \citet{sara+2007}, thus minimizing
  this effect.
To better quantify the separation in $K_s$ between bSGB and fSGB we
first draw by-eye a fiducial line also in the optical-NIR CMD which
follows the mean distribution of the two samples. Then we computed the
difference $\Delta Ks$ between each star, and the magnitude that a
star with the same color would have if lying on this fiducial-line.

The two
histograms shown in Fig.\ref{fig:cmd_seq}, lower panel, illustrate
the results:  we find a difference of 0.20 mag between the two peaks of
the distributions (median values), with $\sigma=0.03$ and $0.05$ for
the upper and lower distributions, respectively.
The distribution is clearly bimodal, 
indicating that 
two distinct populations
are indeed present in NGC~6388.

In other words, the optical--NIR CMD of Fig.\ref{fig:cmd_seq}
reinforces the evidence that the SGB of NGC~6388 shows a split,
similar to that found in NGC~1851.

The KMM test \citep{kmmtest} reveals that a single gaussian fit
  has to be discarded at 93\% level.

\begin{figure*}[htbp]
\begin{center}
\includegraphics[width=0.85\textwidth]{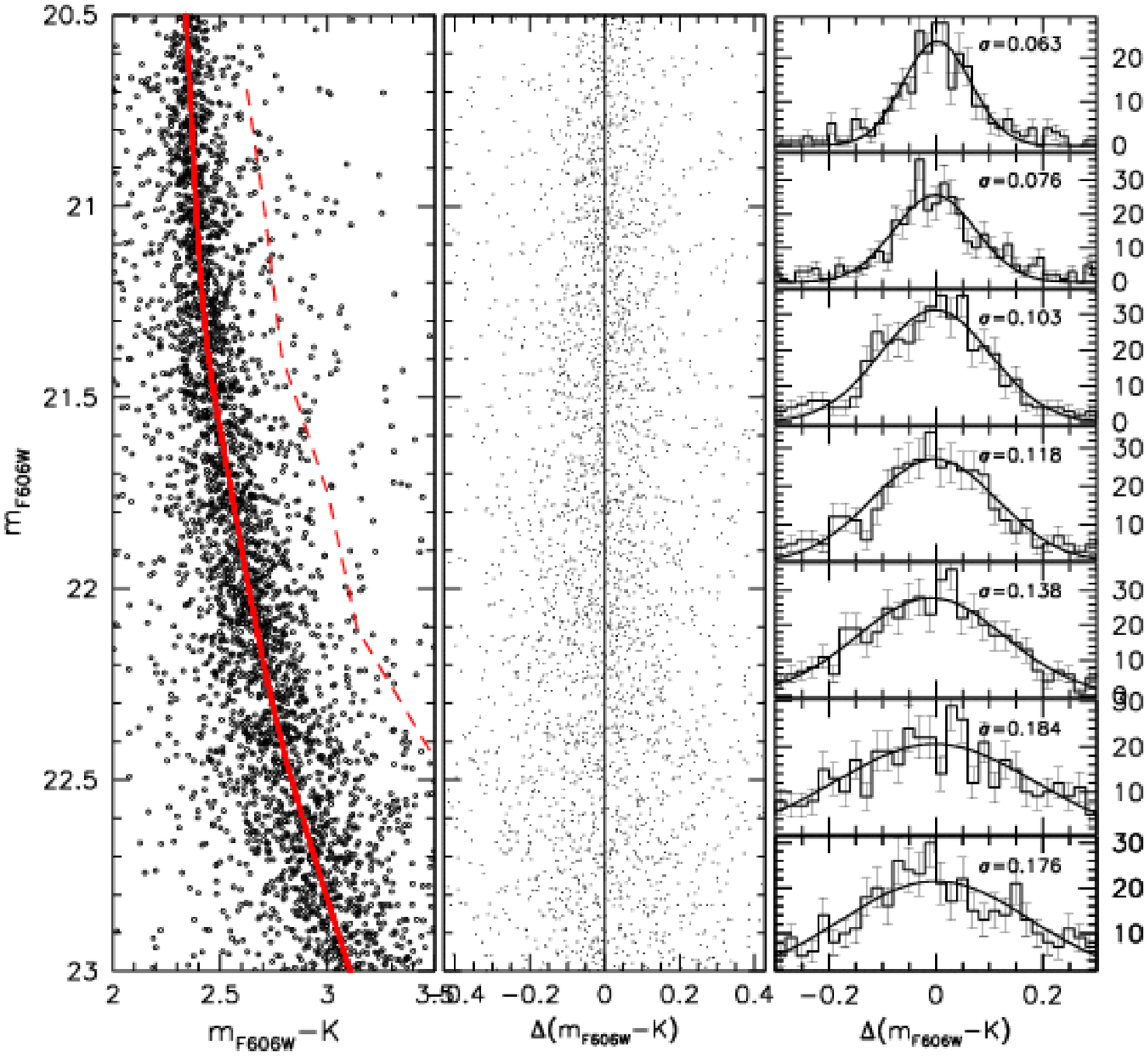}
\caption{Closeup of the main sequence region of the optical-infrared
  diagram. Left panel shows the fiducial line fitting the main
  sequence (drawn by hand); the dashed line represents the one
  $\sigma$ deviation. Middle panel shows the rectified CMD obtained by
  subtracting the color of the fiducial line from each star color;
  right panel shows the histograms of color distribution of stars in
  magnitude bins and the fitting gaussian profiles, together with an
  estimate of the $\sigma$ of the gaussian.}
\label{fig:histo_6388}
\end{center}
\end{figure*}

We also performed a statistical analysis on the main sequence region,
looking for evidence of double sequences.  But the effort produced no
result so far, due to the low precision of the photometry at the
  relatively faint level of the MS.
Following the procedure described in \citet{piot+2007}, we calculated
the color distribution of our selected stars and we fit them
with gaussians.
The left panel of Fig.\ref{fig:histo_6388} shows a close-up of the
optical--NIR CMD. In the middle panel we show the main sequence after
subtraction of a fiducial line drawn by hand (the red continuous line in
the left panel of Fig.\ref{fig:histo_6388}).  The rightmost panel shows 
the color distribution of the stars present in the middle panel, 
grouped into different magnitude bins.  There is no evidence for multiple 
main sequences, a result that we also confirmed by using the KMM 
test \citep{kmmtest}.

\subsection{Central field (NGC~6388-a)}\label{sec:cen}
The central field of NGC~6388
suffers severe crowding.  
The sharp PSF provided by the AO correction makes crowding less of an issue.

\begin{figure*}[htbp]
\begin{center}
\includegraphics[width=0.66\textwidth,angle=270]{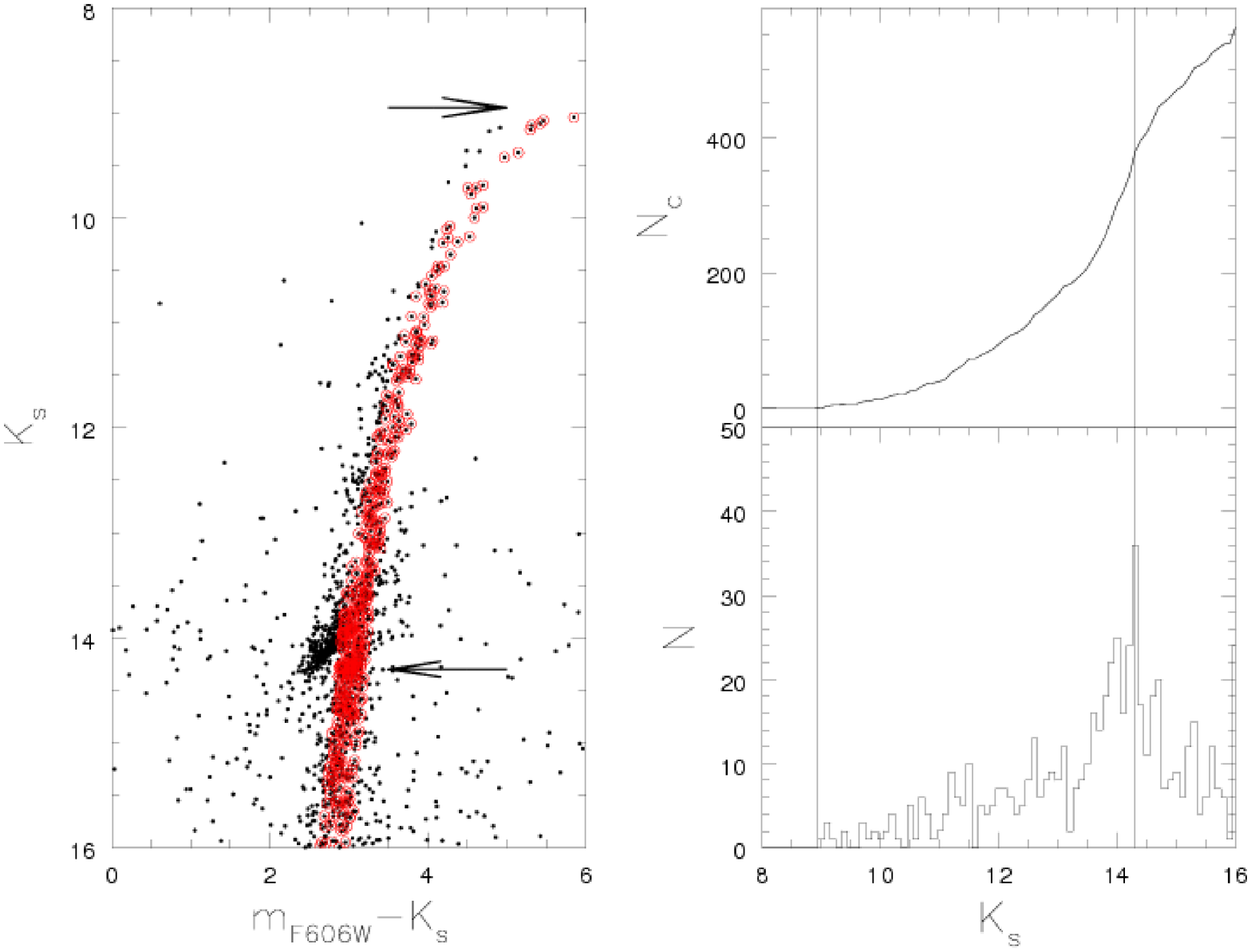}
\caption{Left panel: CMD of NGC~6388-a in $K_s$. Right panel:
  Cumulative (upper right panel) and differential (lower right panel)
  luminosity functions along the RGB. The two arrows show the
  positions of the RGB bump and tip.}
\label{fig:cmd_cen}
\end{center}
\end{figure*}

We cannot directly compare our MAD photometry with other NIR CMDs (such
as those obtained from SOFI or ISAAC), since we have observations in only 
one IR filter (the $K_s$ band).
But we can construct an optical--NIR CMD,
which is shown in Fig.\ref{fig:cmd_cen}.  In order to have the
cleanest possible CMD, we excluded stars located outside a radius of
800 pixels from the center of the image.
To pick the most isolated stars, we also excluded stars with a
DAOPHOT error greater than 0.1.  The number of selected stars is 1975.
Because of the higher crowding, the CMD of the
central field is far from being as deep as the one we obtained for the
external region (see Sec.\ref{sec:ext}), even though the total exposure
time of the analyzed frames is comparable (1000s instead of 1440s).
The left panel of Fig.\ref{fig:cmd_cen} shows the $K_s$ versus 
$m_{\rm F606W}-K_s$ CMD.  Note that only the red clump in the RGB is 
visible: stars hotter (bluer) than the RR Lyrae gap are too faint in 
the 
K-band to be measured in our MAD images.
Red encircled stars in Fig.\ref{fig:cmd_cen} 
have been used to calculate the RGB
luminosity functions (LF), which is shown in the right panels. 
The RGB bump is clearly visible both in the cumulative (upper right panel) 
and in the differential (lower right panel).  The RGB bump and RGB tip
are located at at $K_s=14.30\pm0.05$ and
$K_s=8.95\pm0.05$, respectively.  These values are in
agreement with the ones given in \citet{vale+2007}, taking into
account the errors.  
\section{The age of NGC~6388}

We used our CMDs to estimate the age, reddening, and distance
of NGC~6388 using isochrone fitting.  We used the isochrones by
\citet{piet+2006}, which are calculated for an
$\alpha$--enhanced composition.  Among the available metallicities, we
used the isochrones for [M/H]=-0.35, which corresponds to $Z=0.0080$.
We used the relations given
in \citet{schl+1998} for the $K_s$ extinction coefficient [$A_{K}=0.37
  E(B-V)$], while the absorption coefficient 
in F606W [$A_{\rm F606W}=2.77 E(B-V)$] has been taken from 
\citet{bedi+2005}.

\begin{figure}[htbp]
\begin{center}
\includegraphics[width=0.45\textwidth]{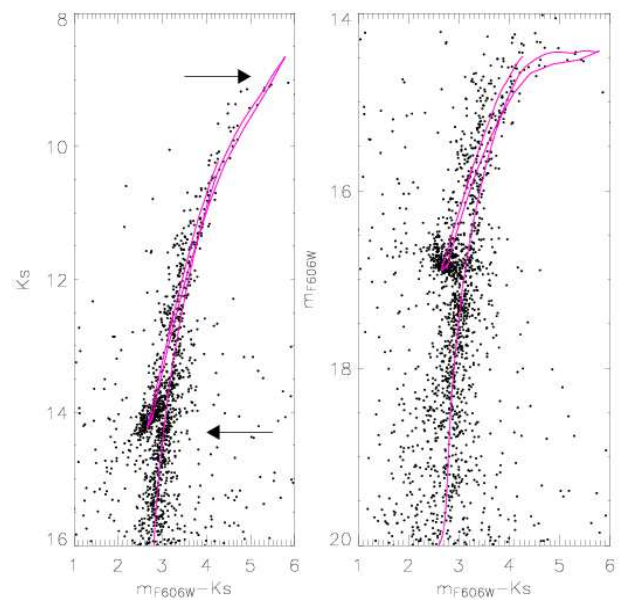}
\caption{Color--magnitude diagram of the central field of NGC~6388 in
  $K_s$ versus $m_{\rm F606W}-K_s$ and $m_{\rm F606W}$ versus $m_{\rm F606W}-K_s$. The
  superimposed isochrone has an age of 11.5 Gyr and a metallicity of
  [M/H]=-0.35. The chemical composition is enhanced in $\alpha$-elements
  following the prescriptions given in \citet{piet+2006}. The best fit
  of the isochrone implies a 
  distance modulus of $(m-M)_\circ=15.33$ and a reddening $E(B-V)=0.38$.}
\label{fig:cmd_iso2}
\end{center}
\end{figure}

\begin{figure}[htbp]
\begin{center}
\includegraphics[width=0.45\textwidth]{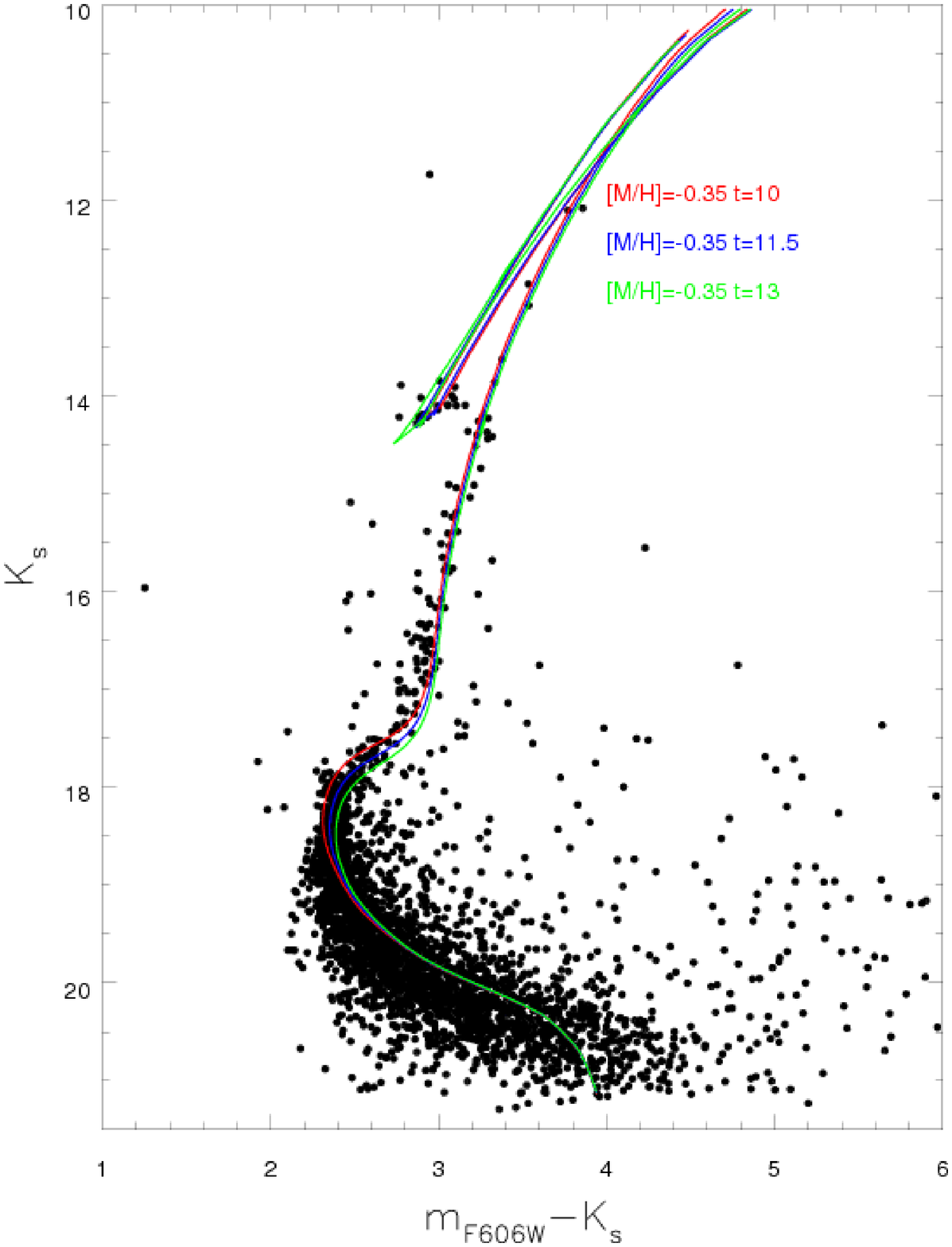}
\includegraphics[width=0.45\textwidth]{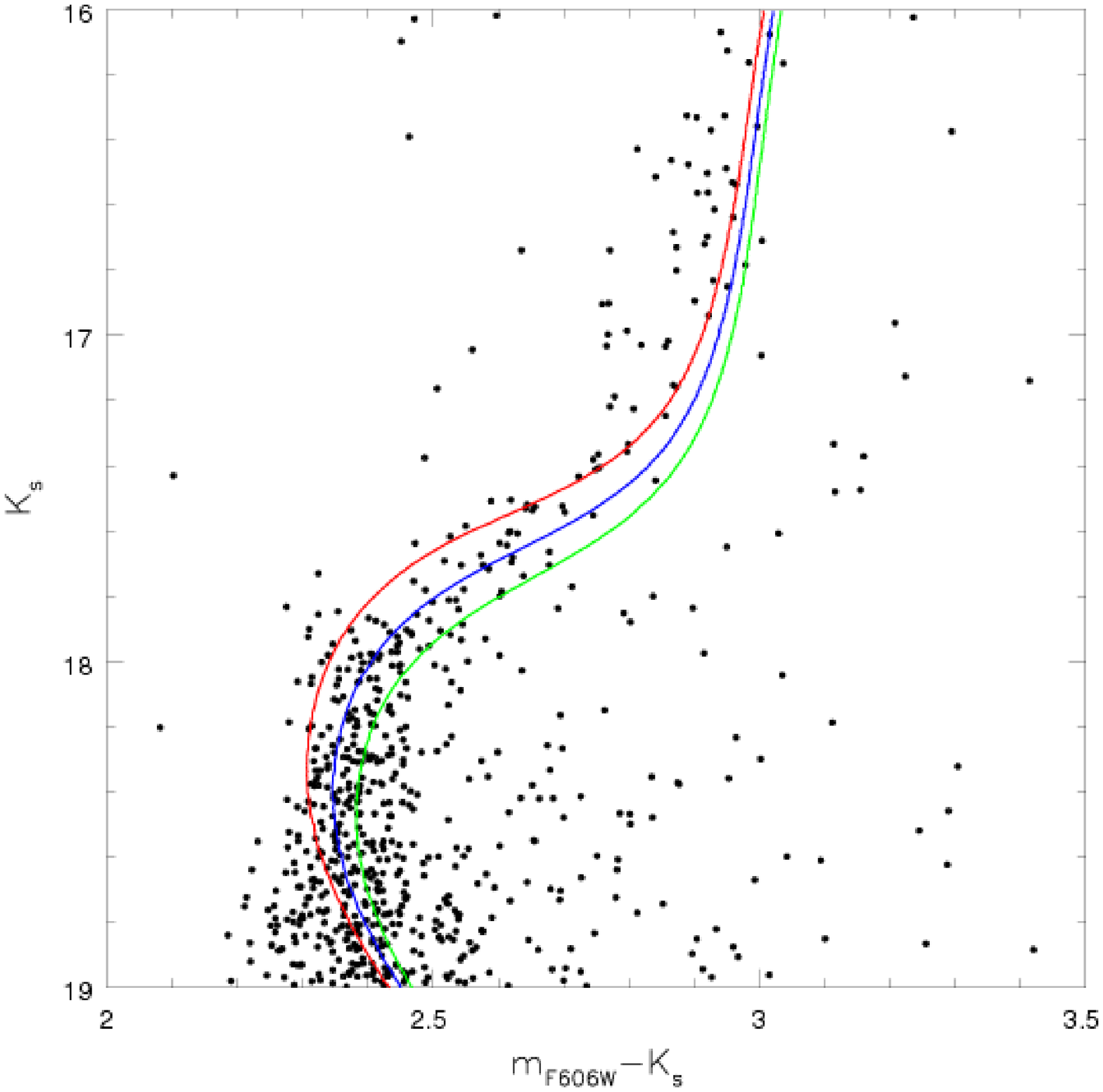}
\caption{Color--magnitude diagrams of the external field of NGC~6388
  (NGC~6388-b).  The superimposed isochrones have ages of 10, 11.5
  and 13 Gyr (red, blue, and green, respectively) and a metallicity of
  [M/H]=-0.35, $\alpha$--enhanced following the prescriptions given in
  \citet{piet+2006}.  The best fit of the HB luminosity and of the
  color of the main sequence stars with 19.0$<K_s<$20.0 implies a 
distance modulus of $(m-M)_\circ=15.33$
  and $E(B-V)=0.44$. The lower panel shows a zoom around the
  SGB region.}
\label{fig:cmd_iso1}
\end{center}
\end{figure}

To begin, we used the 11.5 Gyr isochrone to match the HBs of both the
central and external fields and the MS stars of the external field
with 19.0$<m_{\rm F606W}<$20.0 to fix the distance modulus and
reddening, respectively (Figs.~\ref{fig:cmd_iso2} and
\ref{fig:cmd_iso1}).  In this way, we found $E(B-V)=0.47$ and
$(m-M)_\circ=15.33$.  Note that, in order to fit the RGB of the
central field, we had to use the same distance modulus, but a lower
reddening $E(B-V)=0.38$. Surely, part of the reddening difference
between the two fields is due to differential reddening.  We note that
\citet{hugh+2007} propose an even higher differential reddening than
this.  Still, we cannot exclude that photometric calibration errors
can also contribute to the reddening difference. The distance modulus
and the reddening for the central field are in good agreement with
what was previously found by \citet{vale+2007} and with the values given
in the literature (see for instance \citet{har96} online catalogue,
which gives $E(B-V)=0.37$).

As shown in Fig.~\ref{fig:cmd_iso1}, NGC~6388 has an age of
11.5Gyr$\pm$1.5 Gyr. The large dispersion reflects the dispersion of
the stars along the SGB, as shown in the previous Section. As
discussed in \citet{milo+2008} and \citet{cass+2008} for 
the analogous case of NGC 1851, it is not clear at all that the SGB
dichotomy can be interpreted in terms of an age difference only. A dichotomy in
chemical composition  (e.g. CNO content) could also be behind the
morphology of the SGB in NGC~6388. In that case, the age difference
between the two populations would be greatly reduced
\citep{cass+2008}.

The average age we found is in agreement with many other NGC~6388 age
determinations in the literature.  \citet{cate+2006} used HST
observations and the $Q$ parameter to derive the cluster properties
independently from the reddening and found an age similar to the age
of 47 Tuc, with an error of 1 Gyr \citep[e.g. $\sim11 \pm 1$ Gyr,
see][]{sala+weis2002,grat+2003,dean05}.  This would make our estimate
of the cluster age compatible with the age determined so far.
\citet{hugh+2007} derived an age of $12.6$ Gyr from Washington
photometry, also in agreement with our measurement.  Finally
\citet{yoon+2008} reproduced the CMD of NGC~6388 derived by
\citet{rich+1997} by using two populations of stars both having an age
of $13$ Gyr, but a different helium content (which they choose 
in order to be able to reproduce the blue HB).

\section{Conclusions}\label{sec:discussion}

This paper has presented a pilot study that we carried out in order to
investigate how well a MCAO/NIR system mounted on a 8m telescope can
do photometry in crowded stellar field.
We used the Multi--Conjugated Adaptive Optics Demonstrator (MAD) mounted
on the VLT \citep{marc+2007}, coupled with the Layer-Oriented wavefront
sensor \citep{raga+2000,arci+2006,arci+2007}, to observe in the $K_s$ band
two fields in the highly crowded Bulge GC NGC~6388. The morphology of
the CMD of this cluster still 
represents
an unsolved enigma. We wanted
to verify whether the recently suggested split in its SGB 
(first identified
on ACS/HST images in optical bands) 
can also be seen in the
NIR. Because of the high and differential reddening present in the
cluster, NIR observations are particularly suited for
our purposes.

By combining optical (ACS/HST) data with MAD $K_s$ photometry, we
obtained the deepest optical-nearIR CMD ever produced for NGC~6388.
The good photometric quality at the level of the SGB 
allowed us to construct a $K_s$ vs. $m_{\rm F606W}-K_s$ CMD, which was
previously seen by \citet{piot2008} using only short-baseline visible
observations.
The distribution of the stars along the SGB is clearly bimodal, indicating
the presence of two different stellar populations. Whether this implies
the presence of two groups of stars with a large age difference ($>1$ Gyr), 
or a much smaller age difference but a dichotomy in chemical composition
 (e.g. CNO and Helium) remains to be understood.

In conclusion, this study demonstrates the enormous potential of
the new generation of MCAO instruments that will soon be available on
large ground-based telescopes (LBT, E-ELT, etc.), and the advantage of 
coupling high-resolution optical images from space with ground-based 
AO-corrected NIR data.


\begin{acknowledgements}
Based on observations obtained with the MCAO Demonstrator at the VLT
Melipal Nasmyth focus.  We wish to thank P. Amico for her support
during the observations and S. Cassisi for the set of helium--enriched
isochrones. We also thank the anonymous referee for the careful
reading of the manuscript, and for the useful comments.
\end{acknowledgements}

\bibliographystyle{aa}
\bibliography{biblio}

\end{document}